\documentclass[11pt]{article}
\usepackage{graphicx}
\usepackage{amsfonts, amssymb, amsmath} 
\renewcommand{\H}{{\bf H}}

\newcommand{\tr}{\text{Tr}}

\newcommand{\1}{{\bf 1}}
\newcommand{\ext}{\mbox{ext}}
\newcommand{\V}{{\bf V}}

\newcommand{\one}{{\rm 1\hspace*{-0.4ex}%
\rule{0.1ex}{1.52ex}\hspace*{0.2ex}}}

\title{Four and a Half Axioms for\\  Finite Dimensional Quantum Mechanics} \author{Alexander Wilce\\Department of Mathematical Sciences\\Susquehanna University\\
    Selinsgrove, PA, USA} 

\begin{document}

\maketitle

\begin{abstract} 
I discuss a set of strong, but probabilistically intelligible, axioms from which one can {\em almost} derive the appratus of finite dimensional quantum theory. Stated informally, these require that systems appear completely classical as restricted to a single measurement, that different measurements, and likewise different pure states, be equivalent (up to the action of a compact group of symmetries), and that every state be the marginal of a bipartite 
non-signaling state perfectly correlating two measurements. 

This much yields a mathematical representation of measurements and states that is already very suggestive of quantum mechanics. In particular, in any theory satisfying these axioms, measurements can be represented by orthonormal subsets of, and states, by vectors in, an ordered real Hilbert space -- in the quantum case, the space of Hermitian operators, with its usual tracial inner product. One final postulate (a  simple minimization principle, still in need of a clear interpretation) forces the positive cone of this space to be homogeneous and self-dual and hence, to be the the state space of a formally real Jordan algebra.  From here, the route to the standard framework of 
finite-dimensional quantum mechanics is quite short. \end{abstract} 


\section{Introduction} 
	
It is an old idea that quantum mechanics -- or, a bit more precisely, its probabilistic 
apparatus -- might be derivable from a small number of well-motivated axioms having clear physical, operational or 
probabilistic interpretations.  This goal is all but explicit in 
von Neumann's book \cite{vonNeumann}, and is made both explicit and programmatic in Mackey's
work in the late 1950s \cite{Mackey}. There is a 
small literature of attempts at such a derivation, including the seminal papers of von-Neumann and Birkhoff \cite{Birkhoff-vonNeumann}, 
Zierler \cite{Zierler}, and Piron \cite{Piron}, framing the quantum-logical approach to the problem, and the work of Ludwig \cite{Ludwig}, Gunson \cite{Gunson}, Mielnik \cite{Mielnik}, Araki \cite{Araki} and others, approaching the problem  in terms of ordered linear spaces. More recently, and with a new impetus from quantum information theory, there has been a resurgence of interest in the problem. Whereas most of the earlier work cited above focussed on the structure of single systems, and aimed to obtain the full apparatus of infinite-dimensional (non-relativistic) quantum mechanics, the newer work \cite{BBLW06, BBLW08, Goyal, Hardy, D'Ariano, Rau} has focussed on deriving finite-dimensional quantum mechanics, and has a distinctive emphasis on properties of composite systems. 
Many of these last-cited works (notably those of Hardy \cite{Hardy}, Rau \cite{Rau} and D'Ariano \cite{D'Ariano} 
derive, or come very close to deriving, finite-dimensional QM on the basis of simple and operationally meaningful postulates. 

This paper explores, in a preliminary way, a slightly different -- and possibly less arduous -- route towards such an axiomatization of
finite-dimensional quantum theory, or of something reasonably close
to it. The main ideas are that (i) both classical and quantum systems are
very symmetrical; (ii) irreducible finite-dimensional systems with
homogeneous, self-dual cones are pretty close to being quantum,
thanks to the Koecher-Vinberg Theorem. Therefore, (iii) if we can
somehow use symmetry assumptions to ground homogeneity and
self-duality, we'll be heading in the right direction.

In a bit more detail: the Koecher-Vinberg Theorem \cite{Koecher, Vinberg} classifies 
homogeneous self-dual cones in finite dimensions as the positive 
cones of formally real Jordan algebras. It follows (from the Jordan-von Neumann-Wigner classification of such algebras) that, with the single
exceptional example of the cone of positive $3 \times 3$ matrices
over the octonions, all physical systems having having an irreducible, homogeneous, self-dual cone of 
(un-normalized) states, are either either quantum-mechanical\footnote{allowing here real or quaternionic cases as ``quantum"}, or arise as so-called spin factors, i.e., their normalized state spaces are $n$-dimensional balls. 
Evidently, then, one path the deriving the mathematical framework of QM
from first principles goes by way of supplying an  operational
motivation for homogeneity and self-duality. It will then remain either to
dismiss, or to make room for, spin factors and the exceptional octonionic example as physical models.\footnote{In fact, there is a fairly direct route from Jordan algebras to complex Quantum Mechanics, at least in finite dimensions. A theorem of Hanche-Olsen \cite{Hanche-Olsen} shows that the only Jordan algebras having a 
Jordan-algebraic tensor product with $M_2({\mathbb C})$ -- that is, with a qbit --  are the Jordan parts of $C^{\ast}$-algebras. Since the structure of qbits can be reasonably well-motivated on directly operational grounds, the only irreducible systems in 
a Jordan-algebraic theory supporting a reasonable tensor product, will be full matrix algebras. The condition 
that bipartite states be uniquely determined by the joint probabilities they assign to the two component systems 
-- a condition sometimes called {\em local tomography} -- then dictates that the scalar field be ${\mathbb C}$ \cite{Araki, KRF, Barrett}.} 

Working in a framework in which a physical system is described by specifying a set of basic observables and a finite-dimensional compact, convex set of states \cite{BBLW06, Wilce09b}, I propose four axioms that, taken together, may be glossed as saying that (i) a system appears completely classical as restricted to any single basic observable, (ii) all basic observables are equivalent (up to the action of a compact group of physical symmetries), and (iii) all states arise as marginals of bipartite states that perfectly correlate some pair of basic observables. These assumptions suffice to obtain a representation of measurements as orthonormal subsets of a finite-dimensional ordered real Hilbert space, of states, as vectors therein, and of symmetries, as unitaries acting thereon.  In the quantum case, the space in question is the space of Hermitian operators with its usual tracial inner product. A single additional postulate -- a simple and natural minimization principle, for which one {\em hopes} to find an interpretation -- forces the positive cone of this ordered Hilbert space to be homomgeneous and self-dual with respect to the given inner product.  

Two disclaimers are in order before proceeding, one historical, and the other programmatic. 
First, the general line of attack taken here is not entirely new. The possibility of using the Koecher-Vinberg theorem is mentioned by Gunson \cite{Gunson} as long ago as 1967, but the suggestion seems not to have been followed up (perhaps owing to the then-prevailing focuss on infinite dimensional systems). An exception is the work of Kummer \cite{Kummer}, which, in the context of a quite different set of axioms, also exploits the Koecher-Vingberg Theroem. 
What is novel about the approach taken here is the emphasis on symmetry considerations, and the use of an axiom involving composite systems. 

Secondly, it should be stressed that the postulates discussed below are not
advanced as possible ``laws of thought": the aim here is not to derive QM as 
the {\em uniquely} reasonable non-classical
probability theory (though that would of course be very nice!), but only to 
find simple and transparent {\em characterizations} of quantum probability theory in autonomously
probabilistic or information-theoretic terms, even if still as a theory with contingent elements. 



The balance of this paper is organized as follows. In section 2, I review 
the general framework of abstract probabilistic models elaborated in \cite{BBLW06, BBLW08, Entropy, WilceTA} (and deriving from the older tradition mentioend above, e.g, \cite{Mackey, Ludwig, FR70}). This leads very naturally to a representation of a finite-dimensional physical system in terms of an ordered linear space $V$, generated by the system's normalized states, and its dual space $V^{\ast}$, housing the system's ``effects".  In section 3, I introduce a strong, but physically interpretable, symmetry postulate, namely, that all basic observables, and likewise all pure states, are equivalent up to the action of a compact group. This already 
allows one to introduce an inner product on the space $V^{\ast}$, in terms of which measurements can be represented as orthonormal sets (not necessarily bases) of vectors. With the addition of one purely mathematical regularity requirement (the ``half axiom" of my title) --- a simple extremality condition, satisfied 
in quantum mechanics, and for which one {\em hopes} that an interpretation can be adduced\footnote{As observed by D'Ariano \cite{D'Ariano}, every attempt to date at an axiomatic reconstruction of QM has included at least one such axiom!} --- I obtain the self-duality of the positive cone of $V^{\ast}$.  In section 4, I introduce two additional axioms, each again having a reasonably clear physical meaning, and from these, together with the assumptions of Section 2, deduce that this cone must also be homogeneous.  By the Koecher-Vinberg Theorem, it now follows that $V^{\ast}$ can be represented as a formally real Jordan algebra.  Section 5 collects some final thoughts and poses some questions for further study. Several appendices consider alternative approaches to some of the material in sections 3 and 4.  


\section{Preliminaries}

A {\em test space} \cite{FR70, Wilce09a} is a collection $\mathfrak A$ of
non-empty sets $E, F, ...$, each considered as the outcome-set of
some experiment, measurement, or {\em test}. Subsets of tests are
termed {\em events}. We allow the possibility that distinct tests
may overlap, so that one outcome may belong to several tests. We
write $X = X({\mathfrak A})$ for the total outcome space of $\mathfrak A$,
i.e., $X = \bigcup {\mathfrak A}$. Outcomes $x, y \in X$ are
termed {\em orthogonal}, and we write $x \perp y$, if they are
distinct, but belong to a single test. Note that, at present, there
is no linear structure, let alone an inner product, in view. Note,
too, that we do {\em not} assume that every pairwise-orthogonal set
is an event. A {\em state} on a test space ${\mathfrak A}$ is a mapping
$\alpha : X \rightarrow [0,1]$ such that $\sum_{x \in E} \alpha(x) =
1$ for every test $E \in {\mathfrak A}$. (In other language: a state is
a ``non-contextual" assignment of a probability to every outcome of
every test.)

The simplest case is that in which ${\mathfrak A}$ comprises just a single test, say 
${\mathfrak A} = \{E\}$. In this case, states on ${\mathfrak A}$ are simply probability 
weights on $E$, and we recover discrete classical probability theory.\footnote{Measure-theoretic classical probability theory is also subsumed by this 
framework: if $(S,\Sigma)$ is a measurable space, then the collection 
${\mathfrak B}(S,\Sigma)$ of measurable partitions of $S$ by non-empty 
measurable sets in $\Sigma$ is a test space, and the states on ${\mathfrak B}(S,\Sigma)$ 
are exactly probability measures on $(S,\Sigma)$.} Discrete {\em quantum} probability theory 
arises as the special case in which the test space is the collection ${\mathfrak F}(\H)$ of all {\em frames}, or unordered orthonormal bases, 
of a Hilbert space $\H$. Note that the outcome-set $X$ of ${\mathfrak F}$ is the
unit sphere of $\H$. Gleason's Theorem tells us that states on
$\mathfrak A$ are all of the form $\alpha(x) = \langle \rho x, x \rangle
= \tr(\rho p_x)$ where $\rho$ is a density operator on $\H$ and
$p_x$ is the rank-one orthogonal projection operator $p_x(y) =
\langle y, x \rangle x$.

In many contexts, it is natural to endow the outcome-set $X$ of a test space with a Hausdorff topology for which  
which the relation $\perp$ is closed in $X \times X$ \cite{Wilce09a}. In this case, I shall speak of $X$ as the outcome {\em space} of ${\mathfrak A}$. 
Moreover, I shall assume -- in the main, tacitly -- that all states are continuous in some given topology 
on $X$. This will be important only for the proof of Lemma 1 below.\footnote{It is a consequence of Gleason's Theorem that 
all states on ${\mathfrak F}(\H)$ are continuous in the usual topology on the unit sphere $X$ of $\H$, but in general, 
a test space will admit discontinuous states as well.} 

In what follows, a physical system is modeled by a pair
$({\mathfrak A},\Omega)$, where $\mathfrak A$ is a test space with
outcome-space $X$ and $\Omega$ is a closed, convex, outcome-separating
set of continuous states thereon (possibly a proper subset of the full state space).  
We write $V$ for the cone-base space this
generates (that, is, the linear hull of $\Omega$ in ${\mathbb R}^{X}$,
ordered pointwise on $X$), and $V^{\ast}$ for the order-unit space
dual to $V$. It will be convenient, if a trifle sloppy, to identify
each outcome $x \in X$ with the corresponding evaluation functional,
so that if $\alpha \in \V({\mathfrak A})$, we may write $x(\alpha)$ for
$\alpha(x)$. Writing $u$ for the order unit in $V^{\ast}({\mathfrak
A})$, we have $\sum_{x \in E} x = u$ for every test $E \in {\mathfrak
A}$. Note that if $\mathfrak A$ is the frame manual of a Hilbert space
$\H$, then we have $V^{\ast}({\mathfrak A}) \simeq V({\mathfrak A}) \simeq
{\cal L}_{h}(\H)$, the space of Hermitian operators on $\H$.

If $\mathfrak A$ is any test space, ${\mathfrak A} \times {\mathfrak A}$ denotes the space of {\em product tests} $E \times F$ with $E, F \in {\mathfrak A}$. A state on ${\mathfrak A} \times {\mathfrak A}$ is {\em non-signaling} iff the marginal states $\omega_1(x) = \sum_{y \in F} \omega(x,y)$ and
$\omega_{2}(y) = \sum_{x\in E} \omega(x,y)$ are well-defined, i.e., independent of $E, F \in {\mathfrak A}$. In this case we have conditional states $\omega_{2|x}, \omega_{1|y}$ defined by
\[\omega_{2|x}(y) = \omega(x,y)/\omega_{1}(x) \ \text{and} \ \omega_{1|y}(x) = \omega(x,y)/\omega_{2}(y)\]
(where these make sense, and defined to be $0$ if not). Note that we have laws of total probability: for any test $E \in {\mathfrak A}$,
\begin{equation}\omega_{2}(y) = \sum_{x \in E} \omega_{2|x}(y)\omega_1(x),\end{equation}
and similarly for $\omega_1(y)$. 


In everything that follows, I assume that $V$ is finite-dimensional, and that ${\mathfrak A}$ is {\em locally finite}, meaning that all 
tests in ${\mathfrak A}$ are finite sets. 

\section{Symmetry and Self-Duality} 

In this section, I introduce two operationally transparent axioms, and one simple (but 
much less transparent) minimization condition, and from these deduce that $V^{\ast}_{+}$ is self-dual. 

The first axiom requires that systems be highly symmetrical
in that (i) all outcomes of any given test look alike; (ii) all tests look alike; (iii) all pure states
look alike. This axiom is satisfied by both (discrete) classical and pure quantum systems, and, as we'll see in a moment, already leads to some surprisingly strong consequences.

To make this precise, let us agree that a {\em symmetry} of a system $({\mathfrak A}, \Omega)$ with outcome space $X$ is a homeomorphism $g : X \rightarrow X$ such that (i) $gE \in {\mathfrak A}$ for every test $E \in {\mathfrak A}$, and (ii) $g^{\ast}(\alpha) = \alpha \circ g^{-1}$ belongs to $\Omega$ for every state $\omega \in \Omega$. An action of a group on $({\mathfrak A}, \Omega)$ is an action by symmetries. We say that $\mathfrak A$ is {\em fully symmetric} \cite{Wilce09a} under such an action if (i) all tests have the same cardinality, and (ii) for any bijection $f : E \rightarrow F$ between tests $E, F \in {\mathfrak A}$, there exists some $g \in G$ with $f(x) = gx$ for all $x \in E$. 

Notice that any symmetry $g$ of ${\mathfrak A}$ also determines an affine automorphism of $\Omega({\mathfrak A})$ by $(g\alpha)(x) = \alpha(g^{-1}x)$. We  say that $g$ is a symmetry of the model $({\mathfrak A}, \Omega)$ iff $g\alpha \in \Omega$ for all $\alpha \in \Omega$ and all $g \in G$. It is easy to see that $g$ takes extreme points of $\Omega$ to extreme points of $\Omega$. We shall 
say that an action of $G$ on $({\mathfrak A},\Omega)$ is {\em continuous} iff, for every $\alpha \in \Omega$ and every outcome $x \in X$, 
$g \mapsto \alpha(g^{-1} x)$ is continuous as a function of $G$. 

\begin{quote}{\sf Axiom 1 (Symmetry):  There is a compact group $G$ acting continuously on $({\mathfrak A},\Omega)$, in such a way that 
(i) $G$ acts fully symmetrically on $\mathfrak A$, and (ii) $G$ acts transitively on
$\Omega_{\ext}$.}\end{quote}

A classical test space ${\mathfrak A} = \{E\}$ satisfies Axiom 1 trivially with $G = S(E)$, the symmetric group on $E$. A quantum test space $({\mathfrak F}(\H), \Omega_{\H})$ satisfies Axiom 1 with $G = U(\H)$, the unitary group of $\H$. 


Call an inner product on $V^{\ast}$ {\em positive} iff $\langle a, b
\rangle \geq 0$ for all $a, b \in V^{\ast}_{+}$. Note that the trace
inner product on $V^{\ast} = {\cal L}_{h}(\H)$ is positive in this
sense. 

{\bf Lemma 1:} {\em Subject to Axiom 1, there exists a positive,
$G$-invariant inner product on $V^{\ast}$.}

{\small {\em Proof:} Represent $\Omega_{\ext}$ as $G/K$ where $K =
G_{\alpha}$, the stabilizer of some (any) pure state $\alpha_o$. Any
$f \in V^{\ast}$ gives rise to a function $\hat{f} : G \rightarrow
{\mathbb R}$, defined by $\hat{f}(g) = f(g\alpha_o)$ for all $g \in G$.
This is continuous, so we have an embedding of $V^{\ast}$ as a
$G$-invariant real subspace of the algebra $C(G)$ of continuous
complex-valued functions on $G$. The restriction of the natural
inner product on the latter to $V^{\ast}$ is a real, $G$-invariant
inner product, and is positive, simply because the convolution of
positive functions on $G$ is positive. $\Box$}

Let's agree to call the specific inner product arising from $C(G)$
the {\em canonical} inner product. When we need to differentiate
this from other choices of such an inner product, let's denote it by
$\langle \ , \ \rangle_{G}$.

{\em Remarks:}

(1) The canonical inner product of Lemma 1 satisfies $\langle u, u
\rangle  = 1$, and hence defines a symmetric non-signaling state on
${\mathfrak A} \times {\mathfrak A}$.

(2) Another way to obtain a positive, inner product on $V^{\ast}$ is
just to declare some minimal informationally complete observable orthogonal (as D'Ariano 
notes in [8]). If this observable is $G$-invariant, so will be the 
resulting inner product. 

(3) Appendix D classifies all unitarily invariant positive inner products on
the space $V^{\ast} = {\cal L}_h(\H)$ with $\langle \1, \1 \rangle = 1$. These 
are found to depend on a single real parameter $\lambda \in (0,1]$, with $\lambda = 1$ corresponding 
to the normalized trace inner product. 



{\bf Lemma 2 \cite{Quan-Wilce}:} {\em Let $\langle, \rangle$ be any positive,
$G$-invariant inner product on $V^{\ast}$. There is an embedding $x \mapsto v_x$ of
the outcome-space $X$ into the unit sphere of $V^{\ast}$ with $x \perp y$ implying
$\langle v_x, v_y \rangle = 0$.\footnote{I remind the reader that here, $x \perp y$ means only that 
the outcomes $x, y \in X$ are distinct and belong to a common test; this does not (yet) imply 
that $\langle x, y \rangle = 0$.}}

{\small {\em Proof:} For each $x \in X$, set 
\[q_x = x - \langle x, u \rangle u,\] so that \begin{equation}
\langle q_x, u \rangle = 0.\end{equation} Notice that
$L^{\ast}_\alpha q_x = q_{\alpha x}$ for all $\alpha \in G$ and all
$x \in X$. Since $L$ is unitary and $G$ acts transitively on $X$,
the vectors $q_x$ have a constant norm $\|q_x\| = r$. Moreover,
since $G$ takes any orthogonal pair of outcomes to any other,
$\langle q_x, q_y \rangle$ is constant for any pair $x \perp y$ in
$X$.
Call this value $s_q$. If
$s_q = 0$, we are done: simply set $v_x = q_{x}/\|q_{x}\|$. If
not, we have
\[0 = \langle q_x, 0 \rangle = \langle q_x, \sum_{y \in E} q_y \rangle =
r^2 + (n-1) s_q.\] In particular, $s_q = -\frac{r^2}{n-1} < 0$. In
this case, set $v_x = q_x + cu$ where $c = r/\sqrt{n-1}$ (so that
$s_q = -c^2$). Then, using \theequation, we have $\langle v_{x},
v_{y} \rangle = 0$. Normalizing if necessary, we can take each $v_x$
to be a unit vector. Obviously, the mapping $x \mapsto q_x$ is injective 
$\langle x, u \rangle$ is constant on $X$; hence, so is $x \mapsto v_x$. $\Box$}

In order to get maximum mileage out of this, we impose a very simple, but very strong condition. To set the stage, we need the following observation.

{\bf Lemma 3:} {\em Let $s$ denote the constant value of $\langle x,
y \rangle$ where $x \perp y$. With notation as in Lemma 3, we have,
for all outcomes $x$ and $y$, that
\[\langle v_x, v_y \rangle = \langle x, y \rangle - s.\]}

{\em Proof:} Letting $m$ denote the (constant) value of $\langle x, u \rangle$, we set $q_x = x - mu$ as in the proof of Lemma 3,
so that $\langle q_x, u \rangle = 0$ for all $x$. Recall that $v_x = q_x + cu$ where $-c^2 =s_q$,  the constant value of $\langle q_x, q_y \rangle$ when $x \perp y$. Thus, we have
\[\langle v_x, v_y \rangle  = \langle q_x, q_y \rangle + c^2 \ = \ \langle q_x, q_y \rangle - s_{q}.\]
Now
\[\langle q_x, q_y \rangle = \langle x - mu, y - mu \rangle = \langle x, y \rangle - m \langle x,u \rangle - m \langle u, y \rangle + m^2
= \langle x, y \rangle - m^2.\]
Considering the case where $x \perp y$, this yields
\[s_q = s - m^2.\]
Hence,
\[\langle v_x, v_y \rangle = \langle x, y \rangle - m^2 - s_q  = \langle x, y \rangle - s,\]
as promised. $\Box$

{\bf Definition:} Call a $G$-invariant, positive inner product on $V^{\ast}$ {\em
minimizing} iff the constant $s$ of Lemma 3 is in fact the minimum value of $\langle x, y
\rangle$ on $X \times X$. 

Note that this is certainly the case for the trace inner product on ${\cal L}_h(\H)$, where $s = 0$!

{\bf Lemma 4:} {\em For a minimizing inner product, the vectors $v_x$ of
Lemma 2 lie in the positive cone of $V^{\ast}$.}

{\em Proof:} Immediate from Lemma 3. $\Box$

\begin{quote}{\sf Provisional Axiom 2 (Minimization): There exists a minimizing $G$-invariant, positive
inner product on $V^{\ast}$.}\end{quote}

As we'll see in section 3, {\em all} positive inner products on
$V^{\ast} = {\cal L}_{h}(\H)$ invariant under the unitary group of
$\H$, are in fact minimizing. Thus, it is not out of the question
that Axiom 4 is actually a {\em theorem}. In any case, one would
like to have an operational interpretation for minimization. What
such an interpretation would look like, I'm not sure -- but I'm
betting there is one.

There is one important class of examples in which the existence of a
minimizing inner product {\em does} follow from the previous axioms. Call
a test space $\mathfrak A$ {\em $2$-connected} iff every pair of
outcomes $x, y \in X$ there exist tests $E, F \in {\mathfrak A}$ with $x
\in E, y \in F$ and $E \cap F \not = \emptyset$. Equivalently,
$\mathfrak A$ is $2$-connected iff, for all outcomes $x, y$ there exists
an outcome $z$ with $x \perp z \perp y$. Example: the frame manual
of a Hilbert space.

{\bf Lemma 5:} {\em If $\mathfrak A$ is a fully-symmetric, rank-three,
$2$-connected test space, then any invariant, positive inner product
is minimizing.}

{\small {\em Proof:} Let $x \not \perp y$. By $2$-connectedness, we can find an outcome $z$ with $x \perp z \perp y$. As $\mathfrak A$ has
rank three, we have tests $E = \{x,a,z\}$ and $F  = \{z,b,y\}$. Now, as all outcomes have the same norm in $V^{\ast}$ (here, I conflate an outcome with the corresponding evaluation functional in $V^{\ast}$), we see that $\langle x, t \rangle$ is maximized over outcomes $t$
by $t = x$. Let $s$ be the common value of $\langle e, f \rangle$ where $e$ and $f$ are outcomes with $e \perp f$. Noting that
$x + a + z = u = z + b + y$, we have
\[ \|x\|^2 + 2s = \langle x, u \rangle = s + \langle x, b \rangle + \langle x, y \rangle.\]
This yields $\langle x, b \rangle + \langle x, y \rangle = s + \|x\|^2$. Since $s < \|x\|^2$ and neither $\langle x, b\rangle$ and $\langle x, y \rangle$ can  exceed $\|x\|^2$, it follows that both must exceed $s$.  $\Box$}

{\em Remark:}  2-connectivity seems close to requiring the sets $x^{\perp}$, $x$ ranging over outcomes,
to form a projective geometry. See \cite{Quan-Wilce} for more on this.

{\bf Lemma 6:} {\em Subject to Axiom 1 and Provisional Axiom 2, For every $x \in X$, $\alpha_x(y) := \langle v_x
| v_y \rangle$ defines a state on $\mathfrak A$.}

{\small {\em Proof:} By Lemma 4, $\langle v_x, v_y \rangle \geq 0$ for all $y$. Since $v_x$ and $v_y$ are unit vectors, we also have
$\langle v_x, v_y \rangle \leq 1$ for all $y$. Finally, letting $x \in E \in {\mathfrak A}$, we have, by Lemma 2, and with $v := \sum_{y \in E} v_y$, a multiple of $u$\footnote{since $v_y = q_y + cu$, and $\sum_{y \in E} q_y = 0$, we have $\sum_{y \in E} v_y = ncu$, where $n = |E|$ is independent of $E$ by virtue of $\mathfrak A$'s being fully symmetric}, that
\[\ \ \ \ \langle v_x, v \rangle = \sum_{y \in E} \langle v_x, v_y \rangle = \langle v_x, v_x \rangle = 1. \ \ \Box\]}


We now impose another axiom that, while decidely strong, has a clear 
physical meaning: it says that if we know for certain that a particular outcome will occur, 
then we know the system's state.

\begin{quote}{\sf Axiom 3 (Sharpness): To every outcome $x \in X$,
there corresponds a unique state $\epsilon_x \in \Omega$ with $\epsilon_x
(x) = 1$.}\end{quote}

Note that $\epsilon_x$ is necessarily a pure state. Note, too, that
both (discrete) classical and non-relativistic QM satisfy this
postulate. For some further discussion of (and motivation for) Axiom 3, see Appendix A. 


{\bf Proposition 1:} {\em Subject to Axioms 1-3, $V^{\ast}_{+}$ is self-dual.}

{\small {\em Proof:} Let $\langle, \rangle$ be a minimizing,
$G$-invariant positive inner product. Positivity gives us
$V^{\ast}_{+} \subseteq V^{\ast +} \simeq V_{+}$. Letting $v_x$ be
defined as in Lemma 2, Lemma 7 tells us that $\alpha_{x}(y) : = \langle v_x, v_{y}
\rangle$ defines a state making $x$ certain (since $\langle v_x, v_x
\rangle = \|v_x\| = 1$). By Axiom 3, there is but one such
state, which, by virtue of its uniqueness, is pure. It follows from
Axiom 2 that every pure state has the form $g \alpha_{x} =
\alpha_{gx}$ for some $g \in G$. Thus, every pure state is
represented in the cone $V^{\ast}_{+}$, so that $V^{\ast +}
\subseteq V^{\ast}_{+}$. $\Box$}


An alternative argument, based on slightly different assumptions, is presented in Appendix C.

\section{Correlation, Filtering and Homogeneity}

Having secured the self-duality of $V^{\ast}_{+}$, the next order of business is to secure its homogeneity. This will follow from two 
further axioms. The frst of these tells us that all states of a single system are
consistent with that system's being part of a larger composite in a
state of perfect correlation between some pair of observables. To be more precise,
call a bipartite non-signaling state {\em correlating} iff, for some tests $E, F \in {\mathfrak A}$,
and some bijection $f : E \rightarrow F$, $\omega(xy) = 0$ for all $(x,y) \in E \times F$ with $y \not = f(x)$.

\begin{quote}{\sf Axiom 4 (Correlation):  Every state is the marginal
of a correlating non-signaling state.}\end{quote}

Again, this is satisfied by both classical and quantum systems:
trivially in the first case, and not-so-trivially (i.e., by the
Schmidt decomposition) in the second. 

{\bf Lemma 6 \cite{WilceTA}:} {\em Subject to Axioms 3 and 4,  
for every $\mu \in V_{+}$ there exists a test $E$ such that $\sum_{x
\in E} \mu(x) \epsilon_x$.}

{\small {\em Proof:} Suppose first that $\alpha$ is a normalized state on ${\mathfrak A}$. By Axiom 4, there exists a test space ${\mathfrak B}$ and 
a correlating, non-signaling state $\omega \in \Omega({\mathfrak A} \times {\mathfrak B})$ with $\alpha = \omega_{1}$. Suppose 
$\omega$ correlates $E \in {\mathfrak A}$ with $F \in {\mathfrak B}$ along a bijection $f : E \rightarrow F$. The 
bipartite law of total probability (1) tells us that 
\[\omega_{1}(x) = \sum_{y \in F} \omega_{2}(y) \omega_{1|y} = \sum_{x \in E} \omega_{2}(f(x))\omega_{1|f(x)},\]
where $\omega_{1|y} = \omega_{1|f(x)}$ is the conditional state on $\mathfrak A$ given outcome $y = f(x) \in F$. Since $\omega(x,y) = 0$ for 
$y \not = f(x)$, we have  $\omega_{1|f(x)}(x) = 1$ if $y = f(x) \in F$; thus, $\omega_{2|f(x)} = \epsilon_{x}$, and 
$\alpha = \sum_{x \in E} \alpha(x) \epsilon_{x}$ as promised. 

Now suppose $\mu \in V_{+}$. Then $\mu = r\alpha$ for some $\alpha \in \Omega$ and real constant $r \geq 0$. Expanding $\alpha$ 
as above, we have $\mu = \sum_{x \in E} r\alpha(x)\epsilon_x = \sum_{x \in E} \mu(x) \epsilon_{x}$. 
$\Box$ \\}

The following postulate completes the set.

\begin{quote}
{\sf Axiom 5 (Filtering):  For every test $E$ and every $f : E
\rightarrow (0,1]$, there exists an order-isomorphism $\phi :
V^{\ast} \rightarrow V^{\ast}$ with $\phi(x) = f(x)x$.}\end{quote}

This says that the outcomes of a test can simultaneously and
independently be attenuated by any (non-zero) factors we like by a
reversible physical process. This is equivalent to saying that, for
any test $E$, any outcome $x \in E$, and any $0 < c \leq 1$, there
exists an order-automorphism $\phi$ such that $\phi(x) = cx$ and
$\phi(y) = y$ for all $y \in E \setminus \{x\}$.  This is clearly
the case in both classical and quantum probability theory, and
corresponds to the operationally natural idea that an outcome is
always represented by a physical process, which can be subjected to a
filter reducing its intensity by any specified factor. Possibly one
could make this intuition more precise by considering two-stage
tests (what D'Ariano calls cascades in \cite{D'Ariano}).


{\bf Proposition 2:} {\em Subject to Axioms 1-5, the cone
$V^{\ast}_{+}$ is homogeneous.}\footnote{A different route to homogeneity, via slightly different axioms, is discussed in appendix A.}

{\small {\em Proof:} Let $a$, $b$ be interior points of $V^{\ast}_{+}$. By Proposition 1, $V^{\ast}_{+}$ is self-dual; hence, $\langle a |$ and $\langle b |$ are (un-normalized) states. Let us write $a(x)$ for $\langle a, x \rangle$ and similarly for $b$. By Lemma 6, $\langle a|$ and $\langle b |$ have decompositions $\langle a | = \sum_{x \in E} a(x) \epsilon_x$ and $\langle b | = \sum_{y \in F} b(y) \epsilon_y$ for some pair of tests $E, F \in {\mathfrak A}$.  As $\epsilon_x = \langle x |$, we have $\langle a | = \sum_{x \in E} a(x) \langle x| $, or, more simply, 
$a = \sum_{x \in E} a(x) x$, and similarly for $b$. Since $a$ and $b$ are interior
points, $a(x)$ and $b(y)$ are non-zero for all $x, y$. Let $g$ be a bijection matching $E$ with $F$ (courtesy of Axiom 4), and set
$t(x) = b(gx)/a(x)$. Then, by Axiom 5, there is an order-automorphism $\phi$ of $V^{\ast}$ taking $x$ to $t(x)x$ for every $x \in E$. Hence,  $\phi(a) = \sum_{x \in E} a(x) \phi(x) = \sum_{x \in E} a(x) t(x) x =  \sum_{x \in E} b(gx) x$. Applying $g$, we have
\[\ \ \ g\phi(a) = \sum_{x \in E} b(gx) gx = \sum_{y \in F} b(y) y = b. \ \Box \]}


It now follows from the Koecher-Vinberg theorem that the positive
cone $V({\mathfrak A})_{+}$ is the set of positive elements in a formally real Jordan algebra. 
It is possible that there is a more direct route to this conclusion -- certainly, I 
have not made use of all of the available structure. For example, the
full power of the assumption that $\mathfrak A$ is fully $G$-symmetric (as opposed 
to merely $2$-symmetric) 
is not really exploited. It may also be useful that $V({\mathfrak A})$,
regarded as a subspace of the group algebra ${\mathbb C}[G]$, is closed under convolution,
hence, an algebra -- though the connection between this structure
and the $C^{\ast}$-algebraic structure in QM is not obvious.

Neither is it at all obvious that {\em every} self-dual homogeneous
cone has a representation as $V({\mathfrak A})$ with $\mathfrak A$
satisfying all of the foregoing axioms -- again, full symmetry seems rather
strong, as does Axiom 4 on correlation. It is possible that these axioms constrain
the set of models much more severely.


\section{Summary and open questions}

Axioms 1, 3 and 4 seem natural, or at any rate, {\em intelligible}: one understands what
they {\em say} about a system. Although strong (and certainly, not ``laws of thought"), they 
do identifying a natural class of especially simple and tractable
systems that we might {\em expect} to find well represented ``in
nature". Axiom 5 seems natural in a slightly more restricted
context, in which the measurements we make involve
sending systems through filters that they may or may not pass, with
probabilities that can be attenuated at our discretion. (The idea of a filter 
also shows up prominently in the work of Ludwig \cite{Ludwig} and others following 
in the same path, e.g., \cite{Gunson, Mielnik, Kummer}.) In a broader sense, Axiom 5 captures, at least 
in part, the idea that a system should look {\em completely} classical, as restricted to a single measurement. In particular, a process allowable in classical probability theory should be implementable by a ``physical" process 
acting on $V({\mathfrak A})$. 

Provisional Axiom 2 is obviously more problematic, but on the evidence, seems likely to be
satisfied by a wide range of systems. In order better to understand the scope and significance 
of this postulate, one would like to endow an invariant positive inner
product on $V^{\ast}$ with some operational, perhaps information-theoretic
meaning. 

{\bf Problem 1:} {\em Find an information-theoretic meaning for the canonical 
inner product $\langle \ , \ \rangle_{G}$ arising from the group algebra.} 

In this connection, it would probably be useful to know more about
the non-tracial positive, invariant inner products on ${\cal
L}_{h}(\H)$. In Appendix D, it is shown that the positive, unitarily invariant 
inner products on ${\cal L}_{h}(\H)$ form a one-parameter 
family $\langle a, b \rangle_{\lambda}$, with $0 < \lambda \leq 1$ and 
$\lambda = 1$ corresponding to the trace inner product. It would be 
interesting to know what value of $\lambda$ corresponds
to $\langle \ , \ \rangle_{G}$ where $G$ is the unitary group of
$\H$.

It would of course be nice if something like the result of Appendix D were true in
more generality -- say, for irreducible cones: 

{\bf Problem 2:} {\em Classify the invariant, positive inner
products on (i) any $V^{\ast}$ arising from a model $({\mathfrak A},
\Omega)$ satisfying Axioms 1, 3, 4 and 5; (ii) any irreducible, homogeneous
self-dual cone.}

In this connection, it would also be interesting to know whether some reduction theory is available 
for systems satisfying Axioms 1, 3, 4 and perhaps Axiom 5. Thus, we have 

{\bf Problem 3:} {\em Is a system satisfying Axioms 1, 3, 4 and 5 a direct sum (in some sense!) of irreducible 
such systems? Is the cone of an irreducible such system irreducible as a convex cone?} 

If such a reduction theory is available, there is perhaps a chance that, using the same line of argument as 
that of Appendix C, one may be able to eliminate the troublesome Axiom 2 altogether. 

There are likely many connections between the approach sketched above and that of D'Ariano \cite{D'Ariano}, based on
sequential tests and conditioning operators. (The earlier work of Kummer, cited above, also depends heavily on ideas involving 
sequential measurements and conditioning.) There are probably also connections with the approach of Rau \cite{Rau}, 
which depends upon symmetry considerations similar to, but in some ways stronger than, those of our Axiom 2. 
In particular, Rau assumes that the group $G$ of physical symmetries is a compact Lie group. 
 
{\bf Problem 4:} {\em Clarify these connections. In particular, (i)
clarify the connection between Axiom 5 and the closure of $\mathfrak A$
under formation of compound experiments (``cascades", in the language of \cite{D'Ariano}), together with
the closure of $\Omega$ under conditioning. (ii)  What is the
relationship between the existence of a positive $G$-invariant inner
product and D'Ariano's ``Choi-Jamiolkowski" axiom? \cite{D'Ariano} (iii) What
additional leverage, if any, do we get from the foregoing axioms if $G$ is a
compact Lie group and $X$ is an homogeneous space for $G$?}

Another, particularly important, issue is that of how one can construct, by hand as it were, tensor products
compatible with the foregoing axioms. 

{\bf Problem 5:} {\em Under what conditions do systems satisfying
axioms 1-5, or any subset of these, have non-signalling tensor
products (containing all product states!) that also satisfy these axioms?}

Where this desideratum is met, we would seem to come within hailing distance of Hardy's axioms \cite{Hardy}. See \cite{Wilce09b} for some further discussion of the problems involved in constructing a class of test spaces closed under such a tensor product. 

{\bf Acknowledgement:} I wish to thank Howard Barnum for reading and commenting on an earlier draft of this paper, and, more especially, for introducing me to the papers of Koecher and Vinberg, on which the present exercise depends. Thanks also to C. M. Edwards for pointing out the paper \cite{Hanche-Olsen} of Hanche-Olsen.


\vspace{.2in}

{\bf Appendix A: Entropy and Sharpness} 

The following considerations may offer some independent motivation for Axiom 1. There are two natural ways to extend the definition of entropy to states on a test space. If $\alpha$ is a state on a locally finite, finite-dimensional test space $\mathfrak A$, then Minkowsky's theorem tells us that $\alpha$ has a finite decomposition as a mixture  
$\alpha = \sum_{i} t_i \alpha_i$ of pure states $\alpha_1,...,\alpha_n$. Define the {\em mixing entropy} of $\alpha$, $S(\alpha)$, to be the infimum of $H(t_1,...,t_n) = -\sum_{i} t_i \log(t_i)$ over all such convex decompositions of $\alpha$.  Alternatively, one can consider 
the {\em local entropy} $H_{E}(\alpha) = H(\alpha|_{E}) = -\sum_{x \in E} \alpha(x)\log(\alpha(x))$. Define the {\em measurement entropy} of 
$\alpha$, $H(\alpha)$,to be the infimum value of the local measurement entropies $H_{E}$ over all tests $E$. 

Suppose now that the group $G$ figuring in Axiom 2 is compact. One can then endow ${\mathfrak A}$ with the structure of a compact topological test space \cite{Wilce09a}. Assuming that all states in $\Omega$ are continuous as functions $X \rightarrow {\mathbb R}$, it follows (\cite{Entropy}, Lemma 6) then the infimum defining $H$ is actually achieved, i.e., $H(\alpha) = H_{E}(\alpha)$ for some test $E \in {\mathfrak A}$. 
An easy consequence is that $H(\alpha) = 0$ iff $\alpha(x) = 1$ for some $x \in X({\mathfrak A})$. One can also show \cite{Entropy} 
that $S(\alpha) = 0$ iff $\alpha$ is a limit of pure states. Consequently, if the set of pure states is closed, we have $S(\alpha) = 0$ iff $\alpha$ is pure. 

In both classical and quantum cases, $S = H$. One might consider taking this as a general postulate: 

\begin{quote}{\small {\sf Postulate A: $H(\alpha) = S(\alpha)$ for every state $\alpha \in \Omega$.}} \end{quote}

An immediate consequence is that, subject to the topological assumptions discussed above, a pure state (with mixing entropy $S(\alpha) = 0$) must have local measurement entropy $H_{E}(\alpha) = 0$ for some test $E$, whence, there must be some outcome $x \in E$ with $\alpha(x) = 1$. Conversely, for every $x \in X$, if $\alpha(x) = 1$, then $H_{E}(\alpha) = 0$ for any $E$ containing outcome $x$, whence, $H(\alpha) = 0$. But then $S(\alpha) = 0$ as well, and $\alpha$ is therefore pure. If ${\mathfrak A}$ is {\em unital}, meaning that every outcome has probability $1$ in at least one state, then it follows that ${\mathfrak A}$ is actually sharp.  Moreover, we see that every pure state has the form $\epsilon_x$ for some $x$. In this case, 
the second half of Axiom 2 follows automatically from the first. Further discussion of Postulate A can be found in the paper  \cite{Entropy}, where theories satisfying it are termed {\em monoentropic}. 


{\bf Appendix B: An Alternative Route to Homogeneity}

We say that the space $V$ is {\em weakly self-dual} iff there exists
an {\em order-isomorphism} -- that is, a positive, invertible linear
map with positive inverse -- $\phi : V^{\ast} \rightarrow V$. Note
that such a map corresponds to a positive bilinear form $\omega :
V^{\ast} \times V^{\ast} \rightarrow {\mathbb R}$ (via $\omega(x,y) =
\phi(x)(y)$, hence, to a non-signaling bipartite state on $\mathfrak A$.
We call a bipartite state $\omega$ an {\em isomorphism state} iff
the positive linear map $\hat{\omega} : V^{\ast} \rightarrow V$
given by $\hat{\omega}(x)(y) = \omega(x,y)$ is invertible. One can
show \cite{Steering} that any such state is pure. Note that as $u$ belongs to
the interior of $V^{\ast}_{+}$, if $\omega$ is an isomorphism state,
we must have $\omega_{1} = \hat{\omega}(u)$ in the interior of
$V_{+}$. This suggests the following alternative to Axioms 5:

\begin{quote}{\small {\sf Postulate B: Every interior state is the marginal of an
isomorphism state}}\end{quote}

{\bf Lemma \cite{Steering}:} {\em Subject to Postulate B alone, $V$ is weakly self-dual
and homogeneous.}

{\em Proof:} For there to exist an isomorphism state, $V$ must be
weakly self-dual. For homogeneity, let $\alpha$ and $\beta$ belong
to the interior of $V_{+}$. Then Postulate B implies that there exist
isomorphism states $\omega$ and $\mu$ with $\alpha =
\hat{\omega}(u)$ and $\beta = \hat{\mu}(u)$. Thus, $\beta = (\mu
\circ \omega^{-1})(\alpha)$. As $\mu \circ \omega^{-1}$ is an
order-automorphism of $V$, it follows that the cone is homogeneous.
$\Box$

Postulate B is similar in flavor to Axiom 4, but seems somewhat
awkward in its reference only to states in the interior of $V_{+}$.
It would be desirable to find a single, natural
principle implying both of these axioms. Further work in this direction 
can be found in \cite{Steering}


{\bf Appendix C: An Alternative Route to Self-Duality}





An alternative proof of Proposition 1 (the self-duality of $V_{+}$) appeals to the fact (\cite{Bellisard-Iochum} Lemma 1.0)  that a finite-dimensional ordered space $A$ is self-dual w.r.t a given inner product iff every vector $a \in A$ has a unique Jordan decomposition $a = a_{+} - a_{-}$ with $\langle a_{+}, a_{-} \rangle = 0$. We'll need the following  

{\bf Lemma:} {\em Suppose $A$ carries a positive inner product, with respect to which every element of $A$ has an orthogonal Jordan decomposition. Then $A_{+}$ is self-dual.}   

{\em Proof:} It suffices to show that the orthogonal Jordan decomposition is unique. Suppose $a_{+} - a_{-} = b_{+} - b_{-}$ are two orthogonal Jordan decompositions of an element $a \in A$, and that the inner product is positive. We  $a_{+} - b_{+} = a_{-} - b_{-} =: x \in A$, so that 
\[0 \leq \|x\|^2 = \langle a_{+} - b_{+}, a_{-} - b_{-} \rangle = -(\langle b_{+}, a_{-} \rangle + \langle a_{+}, b_{-} \rangle).\] 
But since the inner product is positive, this last quantity is non-positive: evidently, we must have 
\[\langle a_{+}, b_{-} \rangle = \langle a_{-},b_{+} \rangle = 0,\]
whence, $x = 0$, whence, $a_{+} = b_{+}$ and $a_{-} = b_{-}$: the decomposition is unique, as advertised. $\Box$

{\bf Theorem A:} {\em Suppose $A = V({\mathfrak A})$ is spectral, that $\mathfrak A$ is $2$-symmetric, and that Postulate 3 holds. Then $V({\mathfrak A})$ is self-dual.} 

{\em Proof:} If $f : E \rightarrow {\mathbb R}$, where $E \in {\mathfrak A}$, let $a_f = \sum_{x \in E} f(x) x$. Note that 
this gives us a positive linear mapping ${\mathbb R}^{E} \rightarrow V({\mathfrak A})^{\ast}$. That ${\mathfrak A}$ is spectral implies that every 
{\em positive} element of $V^{\ast}$ has a representation as $a_f$ for some $f \geq 0$ on some $E \in {\mathfrak A}$. Notice that 
$u = a_{1}$ for the constant function $1 : E \rightarrow {\mathbb R}$ on {\em any} test $E \in {\mathfrak A}$. 

Now let $v_x = q_{x} + cu$, where $q_x = x - \langle x, u \rangle u = (1 - \langle x, u \rangle)x$, as in Lemma, so that $v_x \perp v_y$ for 
$x \not = y$ in $E$. If $f \in {\mathbb R}^{E}$, let $v_f = \sum_{x \in E} f(x) v_x$. Note that  
\[v_{f} = \sum_{x \in E} f(x)v_x = \sum_{x \in E} f(x) (1 - \langle x, u \rangle + nc)x.\] Setting 
$g \equiv 1 - \langle x, u \rangle + nc$ (noting that this is constant!), we have $v_{f} = \sum_{x \in E} f(x) v_x = a_{fg}$. In particular, 
$a_{g} = v = ncu$, so that $g \not = 0$. Thus, we have $a_{f} = a_{f/g g} = v_{f/g}$. Thus, if $g \not = 0$, every $a = a_{f}$ in $V^{\ast}$ has an {\em orthogonal} resolution with respect to an orthonormal set $\{v_x | x \in E\}$ for some $E \in {\mathfrak A}$. 
Finally, since (by our provisional Postulate 3) every $v_x \geq 0$, every every vector with an 
orthogonal resolution realtive to the $v_x$ has an orthogonal Jordan decomposition. $\Box$ 




{\bf Appendix D: Invariant positive inner products on ${\cal L}_{h}(\H)$.}

Let $\H$ be a complex Hilbert space of dimension $n$, with frame manual ${\mathfrak F}$ and unit sphere $X$. We seek to classify the unitarily invariant inner products on ${\cal L}_h(\H)$ that are positive on the positive cone of the latter, and to show that all of these are automatically minimizing. 

As remarked above, Gleason's Theorem provides an isomorphism between the space $V({\mathfrak F})$ of signed weights on ${\mathfrak F}$, and the space ${\cal L}_{h}(\H)$ of Hermitian operators on $\H$: for every $\alpha \in V({\mathfrak F})$, there is a unqiue $W_{\alpha} \in {\cal L}_h(\H)$ with $\alpha(x) = \langle W_{\alpha} x , x \rangle$ for all $x \in X$. We also have a dual isomorphism $V^{\ast}({\mathfrak F}) \simeq {\cal L}_h(\H)$, sending each $a \in V^{\ast}({\mathfrak F})$ to an Hermitian operator $A_{a}$ with $\tr(A_a W_{\alpha}) = a(\alpha)$ for all $\alpha \in V({\mathfrak F})$. Note that in this representation, the order unit is represented by the identity operator $\one$ on $\H$.   
If $U$ is a unitary operator on $\H$, understood as acting on $X$, then the natural action on $V({\mathfrak F})$ is given by 
$U(\alpha)(x) = \alpha(U^{-1}x)$ for all $\alpha \in V({\mathfrak F})$ and all $X \in X$. Thus, we have 
$\langle W_{U\alpha}x, x \rangle = \langle W_{\alpha} U^{-1}x, U^{-1}x \rangle$, whence, $W_{U\alpha} = U W_{\alpha}U^{\ast}$ for all 
states $\alpha$. In other words, the natural representation of $U(\H)$ on $V({\mathfrak F}(\H)) \simeq {\cal L}_{h}(\H)$ is exactly its 
usual ajoint action. It follows that the dual action of $U(\H)$ on $V^{\ast}({\mathfrak F})$ is again the adjoint action $A \mapsto U^{\ast} A U$. 
Noting that $\1$ and $\1^{\perp}$, the space of trace-0 Hermitian operators, are both invariant under this action, it follows that the two are 
orthogonal with respect to any unitarily invariant inner product on $V^{\ast}({\mathfrak F})$. Also, since the adjoint representation of $U(\H)$ on $\one^{\perp}$ is irreducible (\cite{VinbergBook}, p. 20), it follows from Schur's Lemma that up to normalization, there is only one unitarily invariant inner product on the latter -- in other words, any invariant inner product on $\one^{\perp}$ has the form $\langle a, b \rangle = \frac{\lambda}{n} \tr(ab)$ for some 
$\lambda > 0$, with $\lambda = 1$ corresponding to the normalized trace inner product.  Hence, an invariant 
inner product on $\V = \langle \one \rangle \oplus \one^{\perp}$ is entirely determined by the normalization of $\one$ and the choice of $\lambda$. 
Taking $\|\one\| = 1$, we have 
that, for any $a = s\one + a_o$ and $b = t\one + b_o$, where $a_o, b_o \in \one^{\perp}$ and $s, t \in {\mathbb R}$, we have 
\[ \langle s1 + a_o, t\one + b_o \rangle = st + \frac{\lambda}{n} \tr(a_o b_o).\] 
We require that $\langle a, b \rangle \geq 0$ for all positive $a, b \in V^{\ast}$. The spectral theorem 
tells us that this is equivalent to requiring that $\langle p_x, p_y \rangle \geq 0$ for all rank-one projections $P_x, P_y$ ($x,y, \in X$). 
Writing  $P_x  = \frac{1}{n}\one + (P_x - \frac{1}{n}\one)$, and similarly for $P_y$, we have  
\begin{eqnarray*}
\langle P_x, P_y \rangle  & = & \frac{1}{n^{2}}  + \lambda \tr \left ( \left ( P_x - \frac{1}{n}\one \right ) \left ( P_y - \frac{1}{n}\one \right ) \right ) \\
& = & \frac{1}{n^2} + \frac{\lambda}{n} \tr \left ( P_x P_y - \frac{P_x + P_y}{n} + \frac{1}{n^2}{\one} \right ) \\
& = & \frac{1}{n^2} + \frac{\lambda}{n} \left ( \tr(P_x P_y) - \frac{2}{n} + \frac{1}{n} \right ) \\
& = & \frac{1}{n^2} + \frac{\lambda}{n} \left ( |\langle x, y \rangle_{o} |^2 - \frac{1}{n} \right ) \\ 
& = & \frac{1 - \lambda}{n^2} + \frac{\lambda}{n} |\langle x, y \rangle_o |^2,\end{eqnarray*}
where $\langle \ , \ \rangle_o$ is the inner product on $\H$. 
This will be non-negative for all choices of unit vectors $x$ and $y$ (in particular, for $x$ and $y$ orthogonal) iff $0 < \lambda \leq 1$ -- in which case, the minimum value of 
$\langle P_x, P_y \rangle$ occurs exactly when $x \perp y$, so such an inner product is automatically minimizing.


\end{document}